\documentclass[12pt]{article}
\usepackage{amssymb,amsmath,epsfig}

\begin{document}

\title{\bf Stability of the Regular Hayward Thin-Shell Wormholes}
\author{M. Sharif \thanks {msharif.math@pu.edu.pk} and Saadia Mumtaz
\thanks{sadiamumtaz17@gmail.com}\\
Department of Mathematics, University of the Punjab,\\
Quaid-e-Azam Campus, Lahore-54590, Pakistan.}

\date{}
\maketitle

\begin{abstract}
The aim of this paper is to construct regular Hayward thin-shell
wormholes and analyze their stability. We adopt Israel formalism to
calculate surface stresses of the shell and check the null and weak
energy conditions for the constructed wormholes. It is found that
the stress-energy tensor components violate the null and weak energy
conditions leading to the presence of exotic matter at the throat.
We analyze the attractive and repulsive characteristics of wormholes
corresponding to $a^r>0$ and $a^r<0$, respectively. We also explore
stability conditions for the existence of traversable thin-shell
wormholes with arbitrarily small amount of fluids describing cosmic
expansion. We find that the spacetime has non-physical regions which
give rise to event horizon for $0<a_{0}<2.8$ and the wormhole
becomes non-traversable producing a black hole. The non-physical
region in the wormhole configuration decreases gradually and
vanishes for the Hayward parameter $l=0.9$. It is concluded that the
Hayward and Van der Waals quintessence parameters increase the
stability of thin-shell wormholes.
\end{abstract}

\section{Introduction}

One of the most interesting attributes of general relativity is the
possible existence of hypothetical geometries having non-trivial
topological structure. Misner and Wheeler \cite{1} described these
topological features of spacetime as solutions of the Einstein field
equations known as wormholes. A ``wormhole" having a tunnel with two
ends allows a short way associating distant regions of the universe.
Besides the lack of some observational evidences, wormholes are
regarded as a part of black holes (BH) family \cite{2}. The simplest
example is the Schwarzschild wormhole that connects one part of the
universe to other through a bridge. This wormhole is not traversable
as it does not allow a two way communication between two regions of
the spacetime leading to the contraction of wormhole throat.

Physicists have been motivated by the proposal of Lorentzian
traversable wormholes given by Morris and Thorne \cite{3}. In case
of traversable wormholes, the wormhole throat is threaded by exotic
matter which causes repulsion against the collapse of wormhole
throat. The most distinguishing property of these wormholes is the
absence of event horizon which enables observers to traverse freely
across the universe. It was shown that a BH solution with horizons
could be converted into wormhole solution by adding some exotic
matter which makes the wormhole stable \cite{3a}. Traversable
wormhole solutions must satisfy the flare-out condition preserving
their geometry due to which the wormhole throat remains open. The
existence of exotic matter yields the violation of null (NEC) and
weak energy conditions (WEC) which is the basic property for
traversable wormholes. Null energy condition is the weakest one
whose violation gives rise to the violation of WEC and strong energy
conditions (SEC). The exotic matter is characterized by
stress-energy tensor components determined through Israel thin-shell
formalism \cite{4}.

Thin-shell wormholes belong to one of the wormhole classes in which
exotic matter is restricted at the hypersurface. To make sure that
the observer does not encounter non-physical zone of BH, a
thin-shell strengthens the wormhole provided that it has an exotic
matter for its maintenance against gravitational collapse. The
physical viability of thin-shell wormholes is a debatable issue due
to inevitable amount of exotic matter which is an essential
ingredient for the existence as well as stability of wormholes. The
amount of exotic matter can be quantified by the volume integral
theorem which is consistent with the concept that a small quantity
of exotic matter is needed to support wormhole \cite{5}. Visser
\cite{6} developed an elegant technique of cut and paste to minimize
the amount of exotic matter by restricting it at the edges of throat
in order to obtain a more viable thin-shell wormhole solution.

It is well-known that thin-shell wormholes are of significant
importance if they are stable. The stable/unstable wormhole models
can be investigated either by applying perturbations or by assuming
equation of state (EoS) supporting exotic matter at the wormhole
throat. In this context, many authors constructed thin-shell
wormholes following Visser's cut and paste procedure and discussed
their stability. Kim and Lee \cite{7} investigated stability of
charged thin-shell wormholes and found that charge affects stability
without affecting the spacetime itself. Ishak and Lake \cite{8}
analyzed stability of spherically symmetric thin-shell wormholes.
Lobo and Crawford \cite{9} studied spherically symmetric thin-shell
wormholes with cosmological constant ($\Lambda$) and found that
stable solutions exist for positive values of $\Lambda$. Eiroa and
Romero \cite{9a} studied linearized stability of charged spherical
thin-shell wormholes and found that presence of charge significantly
increases the possibility of stable wormhole solutions. Sharif and
Azam \cite{10} explored both stable and unstable configurations for
spherical thin-shell wormholes. Sharif and Mumtaz analyzed stable
wormhole solutions from regular ABG \cite{11} and ABGB \cite{12}
spacetimes in the context of different cosmological models for
exotic matter.

It is found that one may construct a traversable wormhole
theoretically with arbitrarily small amount of fluids describing
cosmic expansion. In order to find any realistic source for exotic
matter, different candidates of dark energy have been proposed like
tachyon matter \cite{13}, family of Chaplygin gas \cite{14,14a},
phantom energy \cite{15} and quintessence \cite{15a}. Eiroa
\cite{16} assumed generalized Chaplygin gas to study the dynamics of
spherical thin-shell wormholes. Kuhfittig \cite{17} analyzed
stability of spherical thin-shell wormholes in the presence of
$\Lambda$ and charge by assuming phantom like EoS at the wormhole
throat. Sharif and collaborators discussed stability analysis of
Reissner-Nordstr\"{o}m \cite{18} and Schwarzschild de Sitter as well
as anti-de Sitter \cite{19} thin-shell wormholes in the vicinity of
generalized cosmic Chaplygin gas (GCCG) and modified cosmic
Chaplygin gas (MCCG). Some physical properties of spherical
traversable wormholes \cite{20} as well as stability of cylindrical
thin-shell wormholes \cite{21,21a} have been studied in the context
of GCCG, MCCG and Van der Waals (VDW) quintessence EoS. Recently,
Halilsoy \emph{et al.} \cite{22} discussed stability of thin-shell
wormholes from regular Hayward BH by taking linear, logarithmic and
Chaplygin gas models and found stable solutions for increasing
values of Hayward parameter.

This paper is devoted to construct thin-shell wormholes from regular
Hayward BH by considering three different models of exotic matter at
the throat. The paper is organized as follows. In section
\textbf{2}, we construct regular Hayward thin-shell wormholes and
analyze various physical aspects of these constructed thin-shell
wormholes. Section \textbf{3} deals with stability formalism of the
regular Hayward thin-shell wormholes in the vicinity of VDW
quintessence EoS and Chaplygin gas models. We find different throat
radii numerically and show their expansion or collapse with
different values of parameters. Finally, we provide summary of the
obtained results in the last section.

\section{Regular Hayward Black Hole and Wormhole Construction}

The static spherically symmetric regular Hayward BH \cite{23} is
given by
\begin{equation}\label{1}
ds^2=-F(r)dt^2+F^{-1}(r)dr^2+G(r)(d\theta^2+\sin^2\theta d\phi^2),
\end{equation}
where $G(r)=r^2$ and $F(r)=1-\frac{2Mr^2}{r^3+2Ml^2}$, $M$ and $l$
are positive constants. This regular BH is choosen for thin-shell
wormhole because a regular system can be constructed from a finite
energy and its evolution is more acceptable. This reduces to de
Sitter BH for $r\rightarrow0$, while the metric function for the
Schwarzschild BH is obtained as $r\rightarrow\infty$. Its event
horizon is the largest root of the equation
\begin{equation}
r^3-2Mr^2+2Ml^2=0.
\end{equation}
This analysis of the roots shows a critical ratio
$\frac{l}{M_{*}}=\frac{4}{3\sqrt{3}}$ and radius $r_{*}=\sqrt{3}l$
such that for $r>0$ and $M<M_{*}$, the given spacetime has no event
horizon yielding a regular particle solution. The regular Hayward BH
admits a single horizon if $r=r_{*}$ and  $M=M_{*}$, which
represents a regular extremal BH. At $r=r_{\pm}$ and $M>M_{*}$, the
given spacetime becomes a regular non-extremal BH with two event
horizons.

We implement the standard cut and paste procedure to construct a
timelike thin-shell wormhole. In this context, the interior region
of the regular Hayward BH is cut with $r<a$. The two 4D copies are
obtained which are glued at the hypersurface
$\Sigma^\pm=\Sigma=\{r=a\}$. Infact this technique treats the
hypersurface $\Sigma$ as the minimal surface area called wormhole
throat. The exotic matter is concentrated at the hypersurface making
the wormhole solution a thin-shell. We can take coordinates $\chi^i
=(\tau,\theta,\phi)$ at the shell. The induced 3D metric at $\Sigma$
with throat radius $a=a(\tau)$ is defined as
\begin{equation}\label{4}
ds^2=-d\tau^2+a^2(\tau)(d\theta^2+\sin^2\theta d\phi^2),
\end{equation}
where $\tau$ is the proper time on the shell.

This construction requires the fulfillment of flare-out condition by
the throat radius $a$, i.e., the embedding function $G(r)$ in
Eq.(\ref{1}) should satisfy the relation $G'(a)=2a>0$. The thin
layer of matter on $\Sigma$ causes the extrinsic curvature
discontinuity. In this way, Israel formalism is applied for the
dynamical evolution of thin-shell which allows matching of two
regions of spacetime partitioned by $\Sigma$. We find non-trivial
components of the extrinsic curvature as
\begin{equation}\label{8}
K^\pm_{\tau\tau}=\mp\frac{F'(a)+2\ddot{a}}
{2\sqrt{F(a)+\dot{a}^2}},\quad K^\pm_{\theta\theta}=\pm
a\sqrt{F(a)+\dot{a}^2},\quad
K^\pm_{\phi\phi}=\alpha^2K^\pm_{\theta\theta},
\end{equation}
where dot and prime stand for $\frac{d}{d\tau}$ and $\frac{d}{dr}$,
respectively. To determine surface stresses at the shell, we use
Lanczos equations, which are the Einstein equations given by
\begin{equation}\label{5}
S_{ij}=\frac{1}{8\pi}\{g_{ij}K-[K_{ij}]\},
\end{equation}
where $[K_{ij}]=K_{ij}^+-K_{ij}^-$ and $K=tr[K_{ij}]=[K_{~i}^{i}]$.
The surface energy-momentum tensor $S_{ij}$ yields the surface
energy density $S_{\tau\tau}=\sigma$ and surface pressures
$S_{\theta\theta}=p=S_{\phi\phi}$. Solving Eqs.(\ref{8}) and
(\ref{5}), the surface stresses are obtained as
\begin{eqnarray}\label{9}
\sigma&=&-\frac{1}{2\pi a}\sqrt{F(a)+\dot{a}^2},\\\label{10}
p&=&p_{\theta}=p_{\phi}=\frac{1}{8\pi}\frac{2\dot{a}^2+2a\ddot{a}+2F(a)+
a{F'(a)}}{a\sqrt{F(a)+\dot{a}^2}}.
\end{eqnarray}
\begin{figure}\center
\epsfig{file=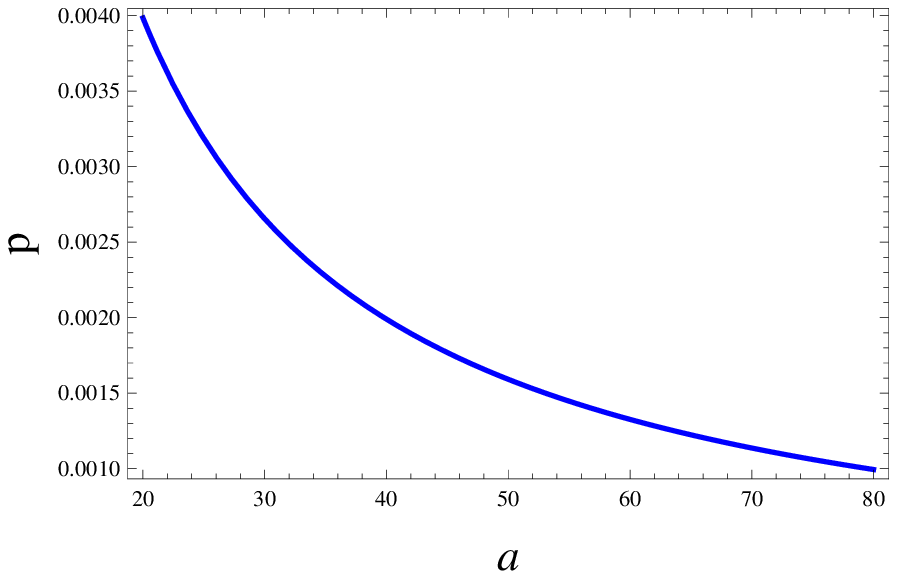,width=0.48\linewidth}
\epsfig{file=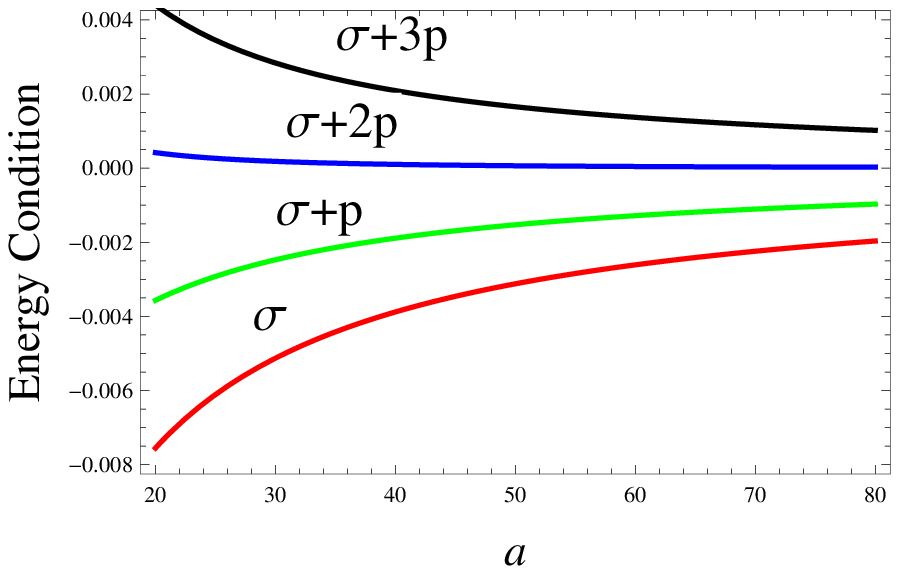,width=0.48\linewidth}\\
\caption{Plots of $p$ and energy conditions with $M=1$ and $l=0.9$.}
\end{figure}
In order to prevent contraction of wormhole throat, matter
distribution of surface energy-momentum tensor must be negative
which indicates the existence of exotic matter making the wormhole
traversable \cite{6}. The amount of this matter should be minimized
for the sake of viable wormhole solutions. We note from
Eqs.(\ref{9}) and (\ref{10}) that $\sigma<0$ and $\sigma+p<0$
showing the violation of NEC and WEC for different values of $M$,
$l$ and $a$. In Figure \textbf{1}, we plot a graph for pressure
showing that pressure is a decreasing function of the throat radius
(left hand side), while the other graph shows violation of energy
conditions associated with regular Hayward thin-shell wormholes.
\begin{figure}\center
\epsfig{file=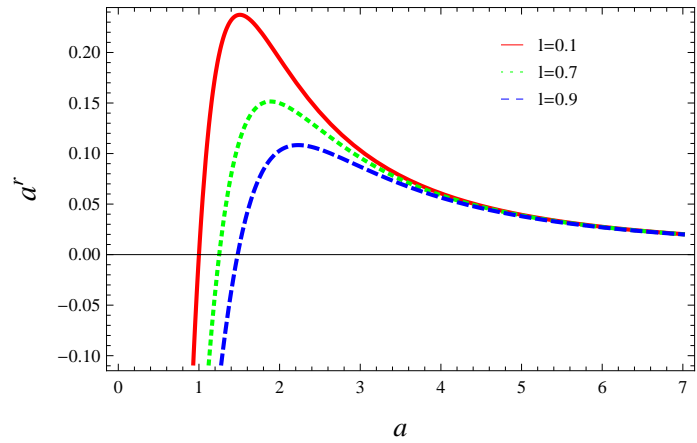,width=0.7\linewidth}\\
\caption{Plots of $a^r$ with $M=1$ and $l=0.1,~0.7,~0.9$. The
wormhole is attractive for $a^r>0$ and repulsive for $a^r<0$.}
\end{figure}

Now we explore the attractive and repulsive characteristics of the
regular Hayward thin-shell wormholes. In this context, we need to
compute the observer's four-acceleration
\begin{equation}\nonumber
a^\mu=u_{;\nu}^\mu u^\nu,
\end{equation}
where $u^\mu=\frac{dx^\mu}{d\tau}=(\frac{1}{\sqrt{F(r)}},0,0,0)$ is
the observer's four-velocity. The non-zero four-acceleration
component corresponding to the given spacetime is calculated as
\begin{equation}\label{12a}
a^r=\Gamma^r_{tt}\left(\frac{dt}{d\tau}\right)^2=\frac{Mr^4-4M^2l^2r}{(r^3+2ml^2)^2},
\end{equation}
for which the geodesic equation has the following form
\begin{equation}\nonumber
\frac{d^2r}{d\tau^2}=-\Gamma^r_{tt}\left(\frac{dt}{d\tau}\right)^2=-a^r.
\end{equation}
An important condition for traversing through wormhole implies that
an observer should not be dragged away by enormous tidal forces. It
is required that the acceleration felt by observer must not exceed
the Earth's acceleration. It is worth stressing here that a wormhole
will be attractive in nature if its radial acceleration is positive,
i.e., $a^r>0$. This supports the fact that an observer must have an
outward-directed radial acceleration $a^r$ in order to keep away
from being pulled by the wormhole. On the other hand, it will
exhibit repulsive characteristics for $a^r<0$. In this case, an
observer must move with an inward directed radial acceleration to
avoid being repelled by the wormhole. The attractive and repulsive
characteristics of the regular Hayward thin-shell wormholes are
shown in Figure \textbf{2}.

Some researchers are excited by the possibility of wormholes in
reality. It appears feasible to keep the wormhole throat open long
enough such that an object can traverse easily through it if the
throat is threaded by exotic matter. The total amount of exotic
matter is quantified by the integral theorem \cite{5}
\begin{equation}
\Omega=\int[\rho+p_{r}]\sqrt{-g} d^3x.
\end{equation}
By introducing radial coordinate $R=r-a$, we have
\begin{equation}
\Omega=\int^{2\pi}_{0}\int^{\pi}_{0}\int^{+\infty}_{-\infty}[\rho+p_{r}]\sqrt{-g}
dR\sin\theta d\theta d\phi.\end{equation}

The wormhole shell, being thin, does not apply any pressure leading
to $p_{r}=0$. Using $\rho=\delta(R)\sigma(a)$, we have
\begin{equation}\label{12b}
\Omega_{a}=\int^{2\pi}_{0}[\rho\sqrt{-g}]|_{r=a}d\phi=2\pi
a\sigma(a).
\end{equation}
Inserting the value of surface energy density $\sigma(a)$, the above
expression yields
\begin{equation}
\Omega_{a}=-\sqrt{\frac{a_{0}^3+2Ml^2-2Ma_{0}^2}{a_{0}^3+2Ml^2}},
\end{equation}
where $a_{0}$ is the wormhole throat radius. It is interesting to
note that construction of a traversable wormhole is possible
theoretically with vanishing amount of exotic matter. This amount
can be made infinitesimally small by choosing exotic fluids
explaining cosmic expansion.

\section{Stability of Thin-Shell Wormholes}

Here we study the formation of thin-shell wormholes from regular
Hayward BH and analyze their stability under linear perturbations.
The surface energy density and surface pressure corresponding to
static wormhole configuration at $a=a_{0}$ become
\begin{equation}\label{12}
\sigma_{0}=-\frac{\sqrt{F(a_{0})}}{2\pi a_{0}}, \quad
p_{0}=\frac{1}{8\pi}\frac{2F(a_{0})+
a_{0}{F'(a_{0})}}{a_{0}\sqrt{F(a_{0})}}.
\end{equation}
The energy density and pressure follow the conservation identity
$S_{~;j}^{ij}=0$, which becomes for the line element (\ref{1}) as
\begin{equation}\label{13}
\frac{d}{d\tau}(\sigma\Phi)+p\frac{d\Phi}{d\tau}=0,
\end{equation}
where $\Phi=4\pi a^2$ corresponds to wormhole throat area. Using
$\sigma'=\frac{\dot{\sigma}}{\dot{a}}$, the above equation can be
written as
\begin{equation}\label{14}
a\sigma'=-2(\sigma+p).
\end{equation}
The thin-shell equation of motion can be obtained by rearranging
Eq.(\ref{9}) as $\dot{a}^2+\Psi(a)=0$, which determines wormhole
dynamics while the potential function $\Psi(a)$ is defined by
\begin{equation}\label{16}
\Psi(a)=F(a)-[2\pi a\sigma(a)]^2.
\end{equation}
In order to explore wormhole stability, we expand $\Psi(a)$ around
$a=a_{0}$ using Taylor's series as
\begin{equation}\label{19}
\Psi(a)=\Psi(a_{0})+{\Psi(a_{0})}(a-a_{0})+\frac{1}{2}\Psi''(a_{0})
(a-a_{0})^2+O[(a-a_{0})^3].
\end{equation}
The first derivtive of Eq.(\ref{16}) through (\ref{14}) takes the
form
\begin{equation}\label{20}
\Psi'(a)=F'(a)+8\pi^2 a\sigma(a)[\sigma(a)+2p(a)].
\end{equation}
The stability of wormhole static solution depends upon
$\Psi''(a_{0})\gtrless0$ and $\Psi'(a_{0})=0=\Psi(a_{0})$. The
surface stresses for static configuration (\ref{12}) yield
\begin{equation}\label{23}
\sigma_{0}=-\frac{\sqrt{a_{0}^3+2Ml^2-2Ma_{0}^2}}{2\pi
a_{0}\sqrt{a_{0}^3+2Ml^2}},\quad
p_{0}=\frac{a_{0}^3+2Ml^2-4Ma_{0}^2}{4\pi
a_{0}\sqrt{(a_{0}^3+2Ml^2)(a_{0}^3+2Ml^2-2Ma_{0}^2)}}.
\end{equation}
The choice of model for exotic matter has significant importance in
the dynamical investigation of thin-shell wormholes. In a recent
work, Halilsoy \emph{et al.} \cite{22} examined the dynamics of
Hayward thin-shell wormholes for linear, logarithmic and Chaplygin
gas models. In this paper, we take VDW quintessence, GCCG and MCCG
fluids at the shell to study stability of regular Hayward thin-shell
wormholes. We will explore the possibility for existence of stable
traversable wormhole solutions by taking different EoS for exotic
matter. In the following, we adopt standard stability formalism by
in the context of the above candidates of dark energy as exotic
matter.

\subsection{Van der Waals Quintessence}

Firstly, we model the exotic matter by VDW quintessence EoS which is
a remarkable scenario to describe accelerated expansion of the
universe without the presence of exotic fluids. The EoS for VDW
quintessence is given by
\begin{equation}\label{11}
p=\frac{\gamma\sigma}{1-B\sigma}-\alpha\sigma^2,
\end{equation}
where $\alpha,~B$ and $\gamma$ are EoS parameters. The specific
values of these parameters lead to accelerated and decelerated
periods. Inserting Eq.(\ref{12}) in (\ref{11}), the equation for
static configuration is obtained as
\begin{eqnarray}\nonumber
&&\left\{a_{0}^2{\psi'(a_{0})}+2a_{0}\psi(a_{0})+\frac{2\alpha}
{\pi}[\psi(a_{0})]^{\frac{3}{2}}\right\}
\left\{2\pi^2a_{0}+B\pi\sqrt{\psi(a_{0})}\right\}\\\label{13}
&&+2\gamma(2\pi a_{0})^2\psi(a_{0})=0.
\end{eqnarray}
The EoS turns out to be
\begin{equation}\label{18a}
\sigma'(a)+2p'(a)=\sigma'(a)\left\{1+\frac{2}{1-B\sigma(a)}
[\gamma-2\alpha\sigma(a)+B\{p(a)+3\alpha\sigma^2(a)\}]\right\}.
\end{equation}
It is found that $\Psi(a)=\Psi'(a)=0$ by substituting the values of
$\sigma(a_{0})$ and $p(a_{0})$, while the second derivative of
$\Psi$ through Eqs.(\ref{20}) and (\ref{18a}) becomes
\begin{eqnarray}\nonumber
\Psi''(a_{0})&=&F''(a_{0})+\frac{[F'(a_{0})]^2}{2F(a_{0})}
\left[\frac{B\sqrt{F(a_{0})}}{2\pi a_{0}+\sqrt{F(a_{0})}}-1\right]
+\frac{F'(a_{0})}{a_{0}}\left[1\right.\\\nonumber
&+&\left.\frac{1}{2\pi a_{0}+\sqrt{F(a_{0})}}\left\{4\pi
a_{0}\gamma+4\alpha\sqrt{F(a_{0})}+B\left(2\sqrt{F(a_{0})}\right.\right.\right.
\\\nonumber&+&\left.\left.\left.\frac{3\alpha B\sqrt{F(a_{0})}}{\pi
a_{0}}\right)\right\}\right]-
\frac{2F(a_{0})}{a_{0}^2}(1+\gamma)\left[1+\frac{1}{2\pi
a_{0}+\sqrt{F(a_{0})}} \right.\\\nonumber&\times& \left.\left \{4\pi
a_{0}\gamma+4\alpha\sqrt{F(a_{0})}+B\left(2\sqrt{F(a_{0})}+\frac{3\alpha
B\sqrt{F(a_{0})}}{\pi a_{0}}\right)\right\}\right].\\\label{22a}
\end{eqnarray}

Now we formulate static solutions for which the dynamical equation
through Eq.(\ref{13}) takes the form
\begin{eqnarray}\nonumber
&&2a_{0}^4(a_{0}^3+Ml^2-Ma_{0}^2)+8M^2l^2a_{0}^2(1-2a_{0}^2)+
\frac{2\alpha}{\pi}\left[2\pi^2a_{0}(a_{0}^3+2Ml^2)
^{\frac{1}{2}}\right.\\\nonumber&&\times\left.
(a_{0}^3+Ml^2-2Ma_{0}^2)^{\frac{3}{2}}+B\pi
(a_{0}^3+2Ml^2-2Ma_{0}^2)^2\right]+2\gamma(2\pi
a_{0}^2)^2\\\label{24a}&&\times[a_{0}^3(a_{0}^3+4Ml^2-2Ma_{0}^2)
+4Ml^2a_{0}^2(a_{0}-1)]=0,
\end{eqnarray}
whose solutions correspond to static Hayward thin-shell wormholes.
In order to explore the wormhole stability, we evaluate numerical
value of throat radius $a_{0}$ (static) from Eq.(\ref{24a}) and
substitute in Eq.(\ref{22a}). We choose Hayward parameter
$l=0,0.1,0.7,0.9$ and check the role of increasing values of $l$ on
the stability of Hayward thin-shell wormholes. We are interested to
find the possibility for existence of traversable thin-shell
wormholes and to check whether the wormhole throat will expand or
collapse under perturbation. For the existence of static stable
solutions, $\Psi''>0$ and $a_{0}>r_{h}$, while $\Psi''<0$ and
$a_{0}>r_{h}$ hold for unstable solutions. For $a_{0}\leq r_{h}$, no
static solution exists leading to non-physical region (grey zone).
In this region, the stress-energy tensor may vanish leading to an
event horizon which makes the wormhole no more traversable. The
stable and unstable solutions correspond to green and yellow zones,
respectively. The graphical results in Figures \textbf{3}-\textbf{4}
can be summarized as follows.

For $\gamma \in (-\infty,-0.3]$, only unstable solutions exist
corresponding to $l=0,0.1$, while both (stable and unstable)
wormhole configurations appear by increasing the values of Hayward
parameter, i.e., $l=0.7,0.9$ as shown in Figure \textbf{3}. For
$l=0.7$, the wormhole is initially stable but its throat continues
to expand leading to unstable solution. The non-physical region in
the wormhole configuration decreases gradually and vanishes for
$l=0.9$. In this case, we find unstable wormhole solution for
$a_{0}<2$ which leads to the collapse of wormhole throat as
$B\alpha^{-(1+\gamma)}$ approaches its maximum value. For increasing
$\gamma$, i.e., $\gamma \in [0.1,0.9]$, there exist unstable and
stable configurations.

Finally, we examine the stability of Hayward thin-shell wormholes
for $\gamma \in [1,\infty)$ and find only stable solutions with
$l=0,0.1,0.7$ which shows the expansion and traversability of
wormhole throat. The unstable solution also appears for $l=0.9$ and
$a_{0}<2$ leading to non-traversable wormhole due to its collapse.
We find stable solutions for $a_{0}>2$ and the wormhole throat
expands which let the wormhole to open its mouth. Figure \textbf{4}
shows the corresponding results for $\gamma=1$. This graphical
analysis shows that the wormhole exhibits physical regions
(stable/unstable) corresponding to different values of Hayward
parameter. Since the regular Hayward wormholes are singularity free
due to their regular centers but the spacetime has event horizons
which give rise to non-physical regions for $0<a_{0}<2.8$ and makes
the wormhole non-traversable. We find that the wormhole can be made
traversable as well as stable by tuning the Hayward parameter to its
large value. Also, it is noted that $\gamma=1$ is the most fitted
value to analyze only stable solutions.

\begin{figure}\center
\epsfig{file=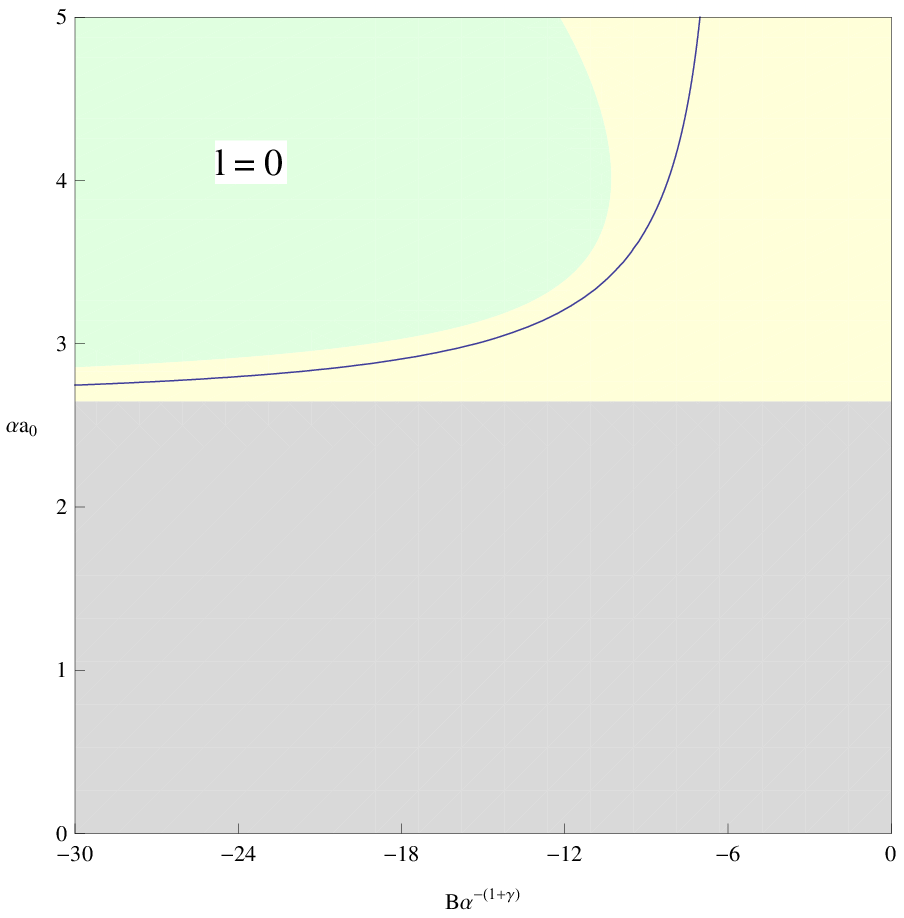,width=0.55\linewidth}\epsfig{file=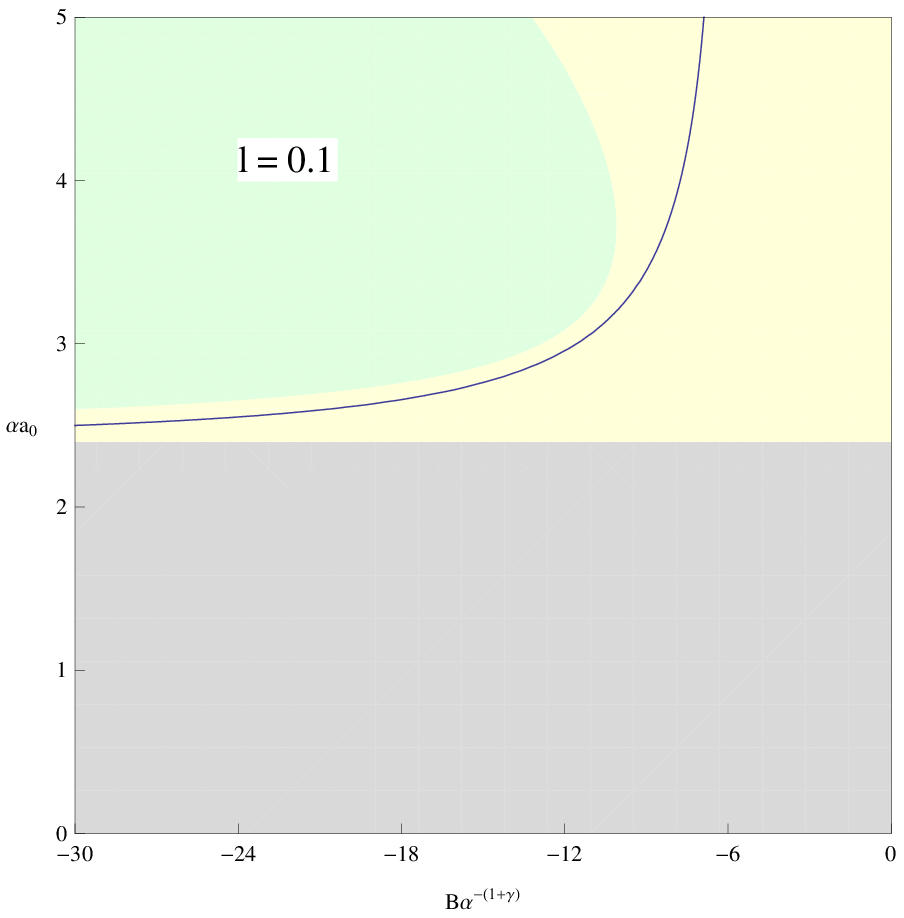,
width=0.55\linewidth}\\
\epsfig{file=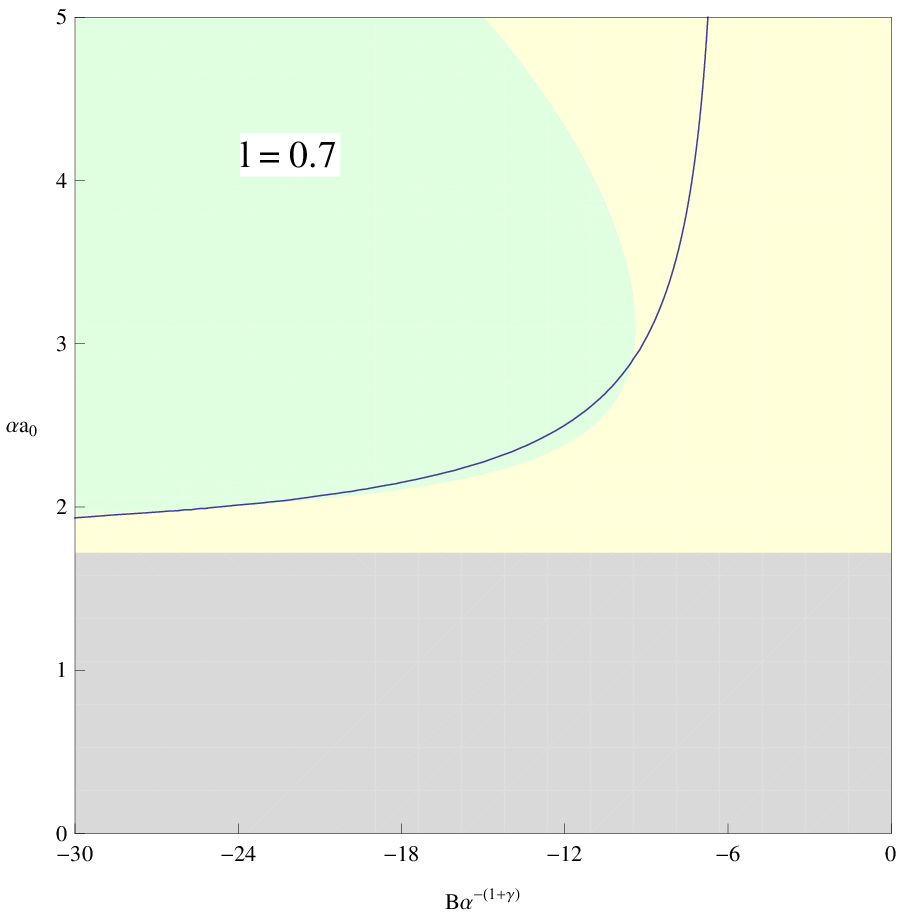,width=0.55\linewidth}\epsfig{file=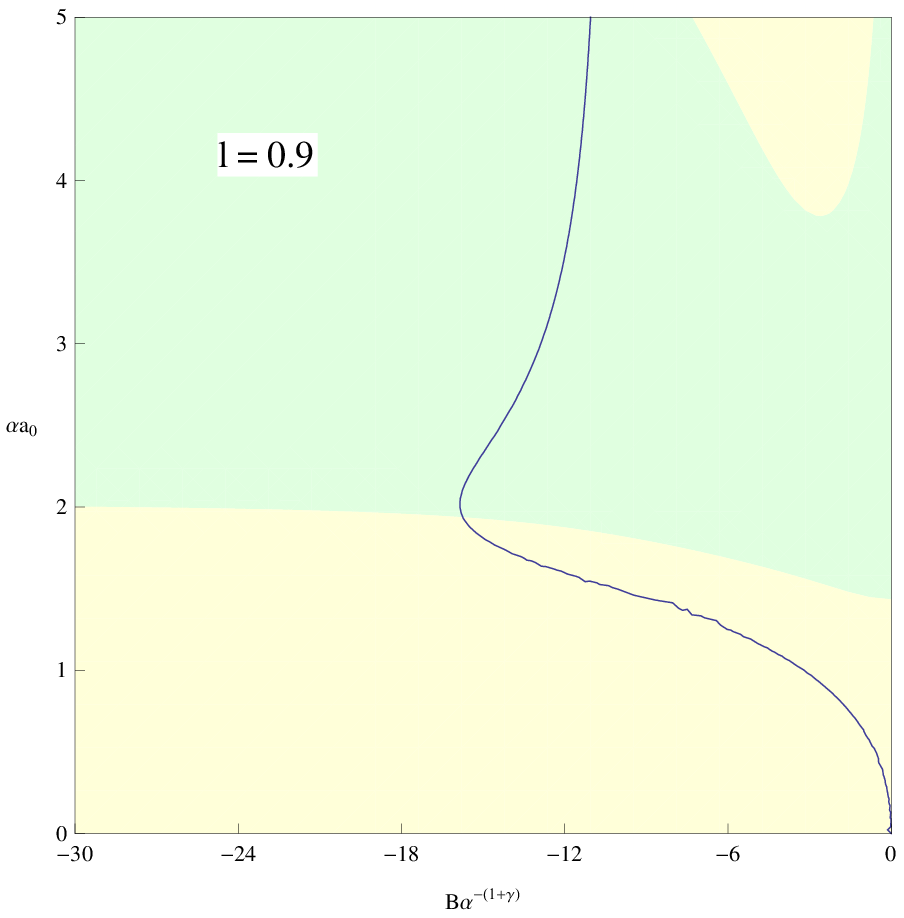,
width=0.55\linewidth}\caption{Plots for regular Hayward thin-shell
wormholes by taking VDW quintessence with $\gamma=-0.5, ~M=1,
~\alpha=1$ and different values of Hayward parameter $l$. The stable
and unstable regions are represented by green and yellow colors,
respectively, while the grey zone corresponds to non-physical
region. Here $B\alpha^{-(1+\gamma)}$ and $\alpha a_{0}$ are labeled
along abscissa and ordinate, respectively.}
\end{figure}
\begin{figure}\center
\epsfig{file=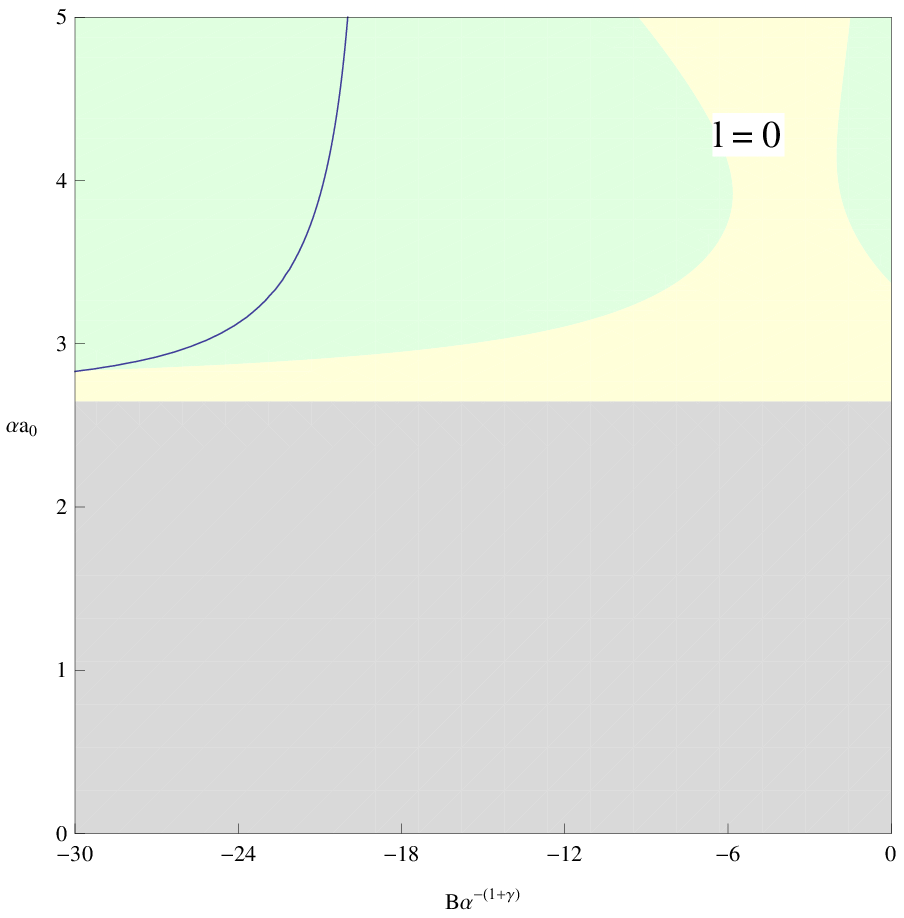,width=0.55\linewidth}\epsfig{file=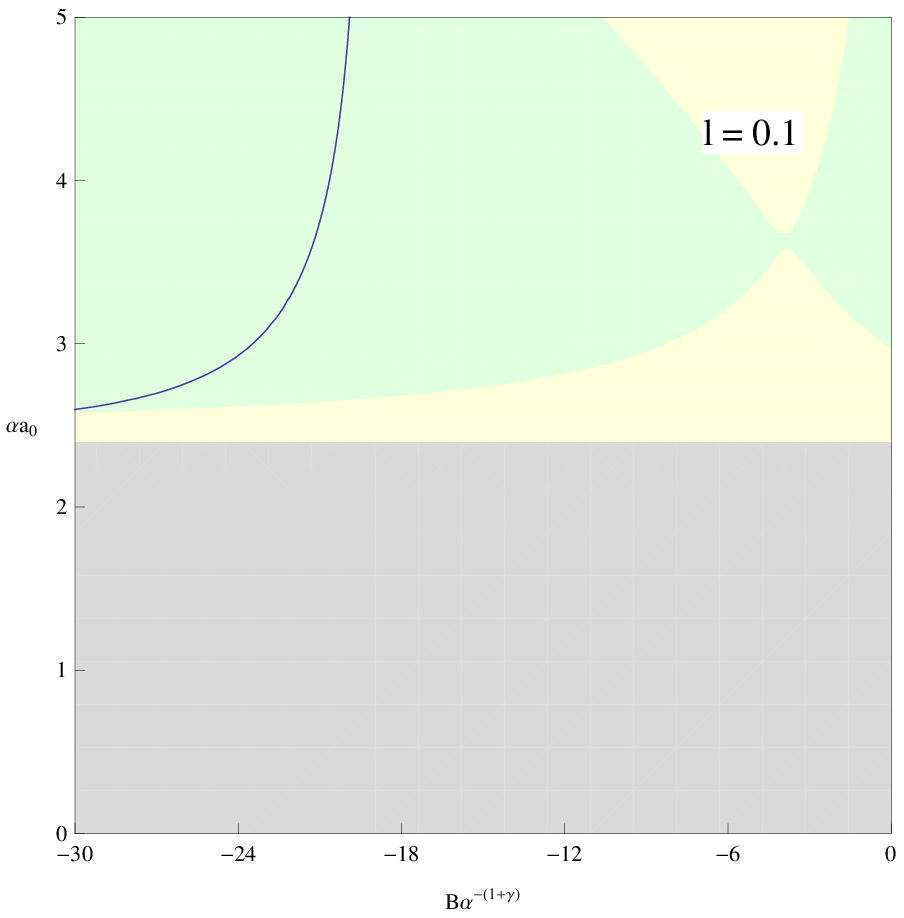,
width=0.55\linewidth}\\
\epsfig{file=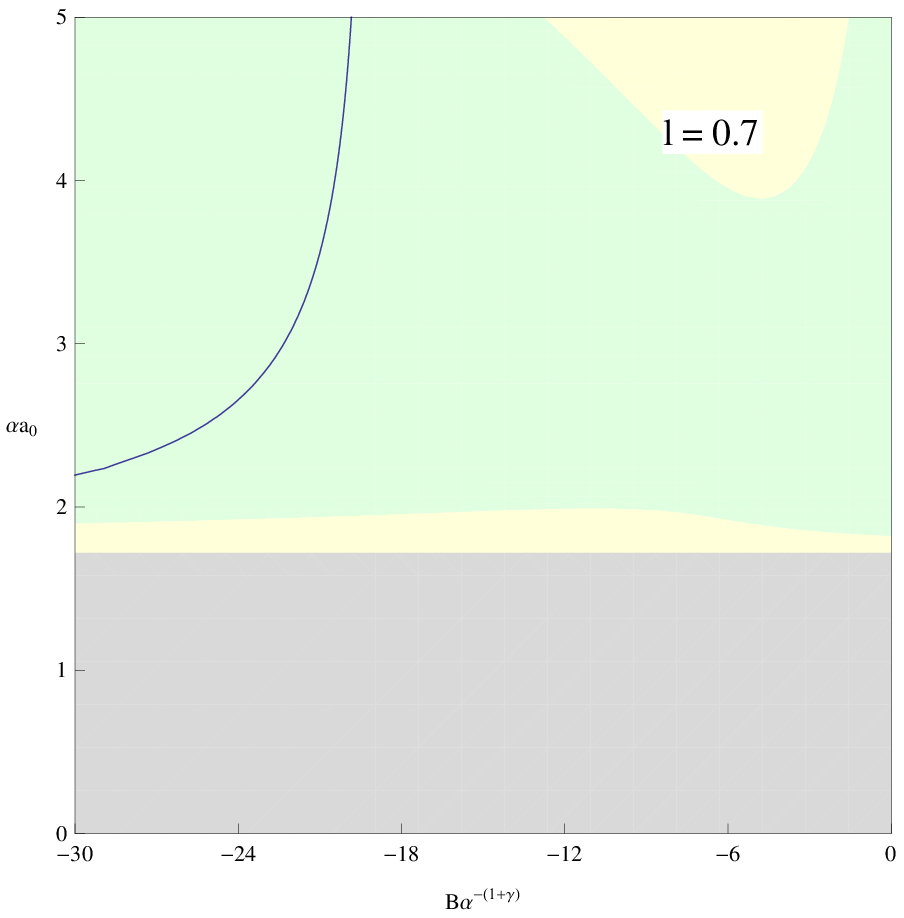,width=0.55\linewidth}\epsfig{file=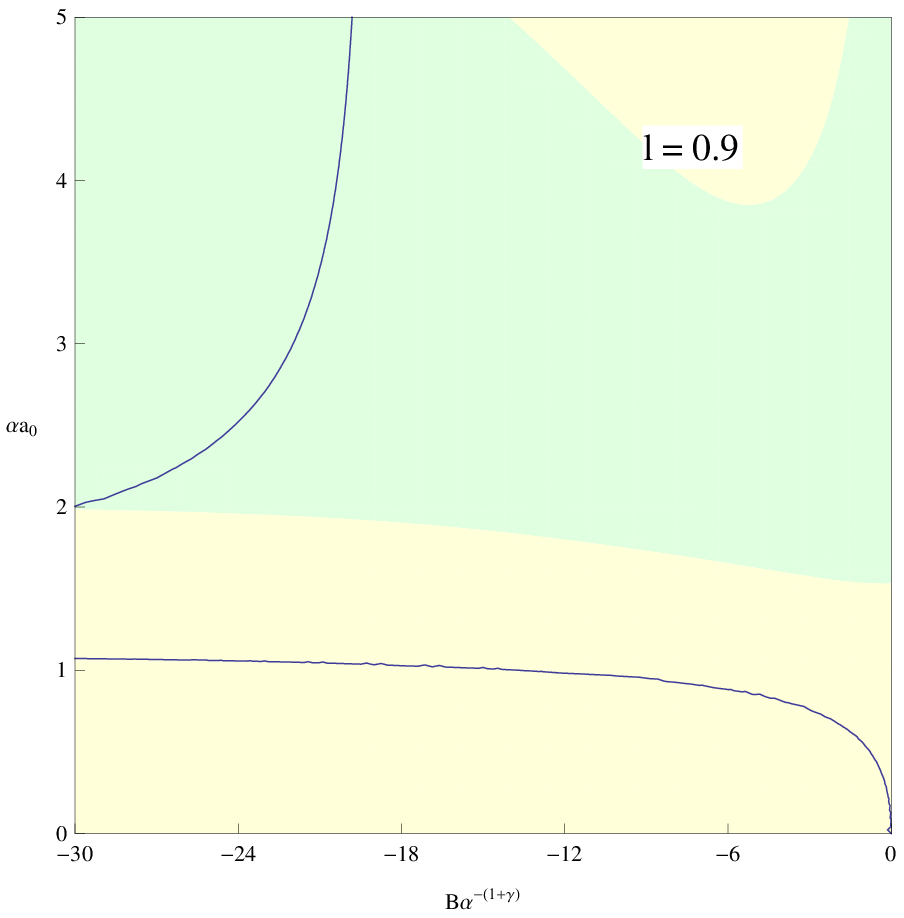,
width=0.55\linewidth}\caption{Plots for VDW quintessence EoS taking
$\gamma=1,~M=1,~\alpha=1$ and different values of Hayward parameter
$l$.}
\end{figure}

\subsection{Generalized Cosmic Chaplygin Gas}

Now we assume GCCG \cite{24} to support the exotic matter at the
shell. Chaplygin gas is a hypothetical substance that satisfies an
exotic EoS. The EoS for GCCG is defined as
\begin{equation}\label{11a}
p=-\frac{1}{\sigma^\gamma}\left[E+(\sigma^{1+\gamma}-E)^{-w}\right],
\end{equation}
where $E=\frac{B}{1+w}-1,~B\in(-\infty,\infty)$, $-C<w<0$ and $C$ is
a positive constant rather than unity. The dynamical equation for
static solutions through Eqs.(\ref{12}) and (\ref{11a}) yield
\begin{eqnarray}\nonumber
&&[a^2_{0}F'(a_{0})+2a_{0}F(a_{0})][2a_{0}]^\gamma-2(4\pi
a^2_{0})^{1+\gamma}[F(a_{0})]^{\frac{1-\gamma}{2}}
\\\label{30}&&\times\left[E+\left\{(2\pi
a_{0})^{-(1+\gamma)}(F(a_{0}))^{\frac{(1+\gamma)}{2}}-E\right\}^{-w}\right]
=0.
\end{eqnarray}
Differentiation of Eq.(\ref{11a}) with respect to $a$ leads to
\begin{equation}\label{18}
\sigma'(a)+2p'(a)=\sigma'(a)\left[1+2w(1+\gamma)\{\sigma^{1+\gamma}-E\}^
{-(1+w)}-\frac{2\gamma p(a)}{\sigma(a)}\right],
\end{equation}
which determines the second derivative of potential function as
\begin{eqnarray}\nonumber
\Psi''(a_{0})&=&F''(a_{0})+\frac{(\gamma-1)[F'(a_{0})]^2}{2F(a_{0})}
+\frac{F'(a_{0})}{a_{0}}\left[1+2w(1+\gamma)\right.\\\nonumber
&\times&\left.\left\{\left(\frac{\sqrt{F(a_{0})}}{2\pi
a_{0}}\right)^{1+\gamma}-E\right\}^{-(1+w)}\right]-
\frac{2F(a_{0})}{a_{0}^2}(1+\gamma)\\\label{22}&\times&\left[1+2w\left
\{\left(\frac{\sqrt{F(a_{0})}}{2\pi
a_{0}}\right)^{1+\gamma}-E\right\}^{-(1+w)}\right].
\end{eqnarray}
Using Eq.(\ref{23}) in (\ref{30}), the corresponding dynamical
equation for static solution becomes
\begin{eqnarray}\nonumber
&&a_{0}^3+2Ml^2-4Ma_{0}^2-2(2\pi
a_{0})^{1+\gamma}[a_{0}^3+2Ml^2-2Ma_{0}^2]^{\frac{1-\gamma}{2}}
\left[E\right.\\\nonumber&&+\left.\{(2\pi a_{0})^{-(1+\gamma)}
(a_{0}^3+2Ml^2)^{\frac{-(1+\gamma)}{2}}(a_{0}^3+2Ml^2-2Ma_{0}^2)
^{\frac{(1+\gamma)}{2}}-E\}^{-w}\right]=0,\\\label{24}
\end{eqnarray}
which gives static regular Hayward wormhole solutions. Here we again
employ the same technique for the stability analysis as in the
previous subsection. The results in Figures \textbf{5}-\textbf{6}
correspond to GCCG. For $\gamma=0.2,1$ and $l=0,0.1,0.7$, we find
fluctuating wormhole solutions. It is found that unstable solution
exists for small values of throat radius $a_{0}$, while the
solutions become stable when the throat radius expands. There exists
a non-physical region for $0<a_{0}<2$ which diminishes with $l=0.9$
making a stable traversable wormhole. For $\gamma=1$ and $l=0.9$,
there is a fluctuating behavior of wormhole throat. Initially, it is
stable for smaller values of $a_{0}$ but becomes unstable by
increasing throat radius. Finally, we analyze stable regular Hayward
thin-shell wormhole for throat radius $a_{0}>2$ which undergoes
expansion.
\begin{figure}\center
\epsfig{file=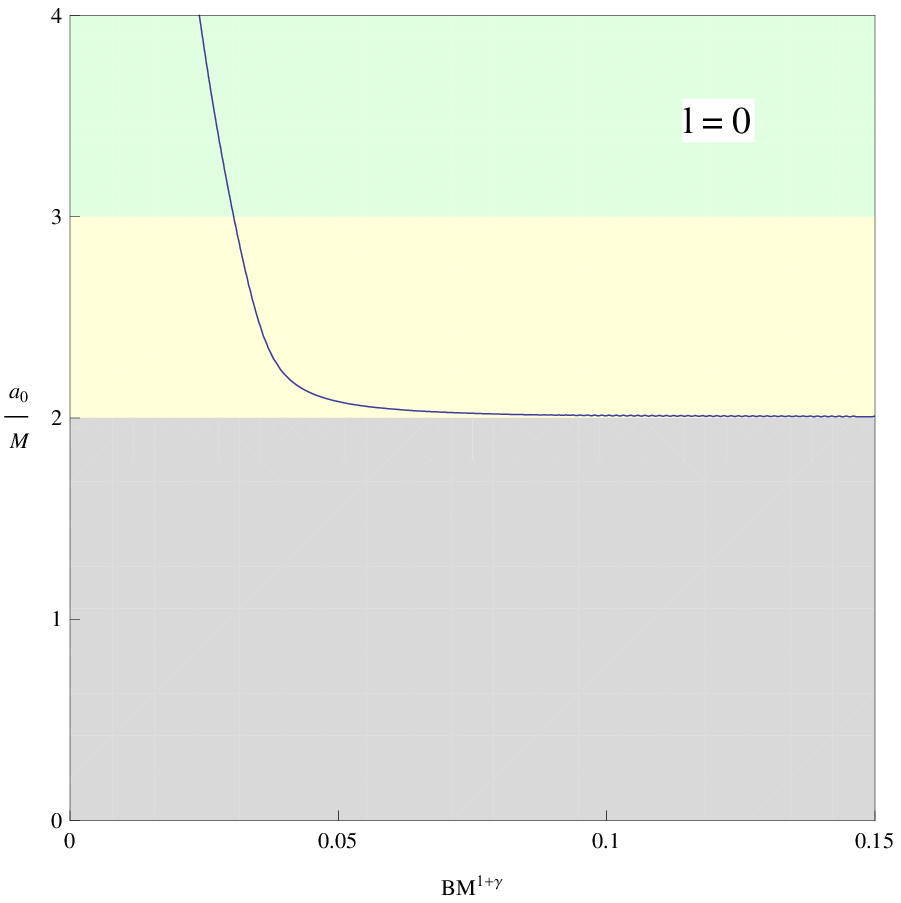,width=0.55\linewidth}\epsfig{file=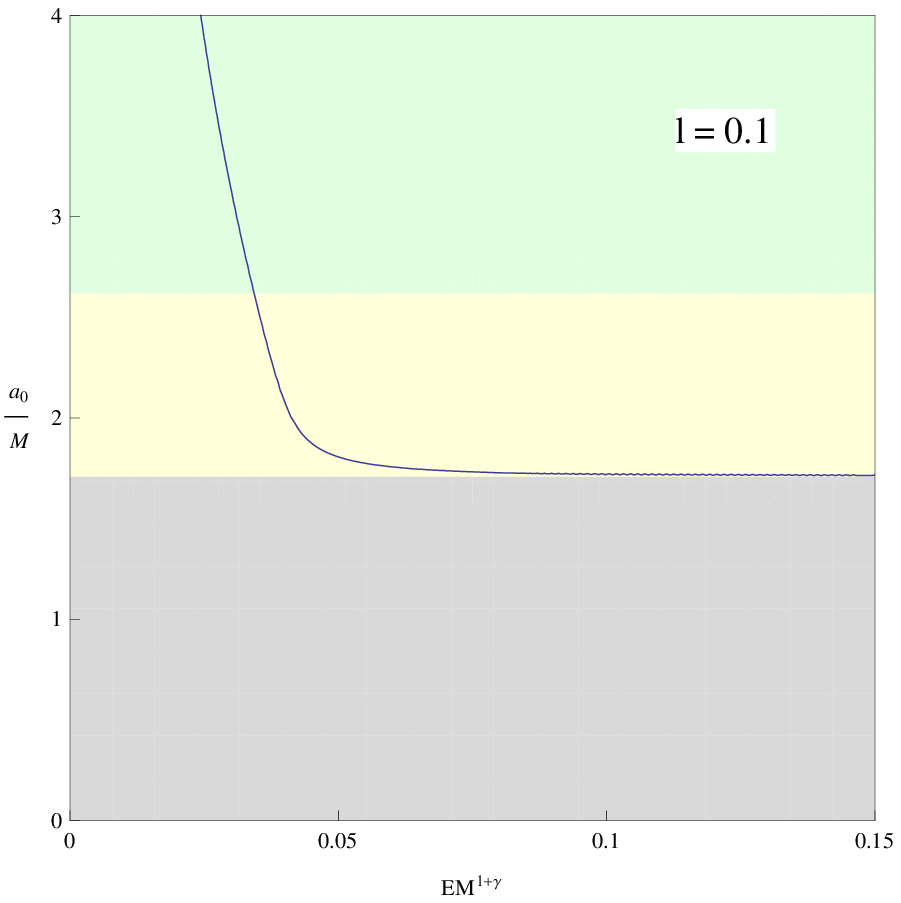,
width=0.55\linewidth}\\
\epsfig{file=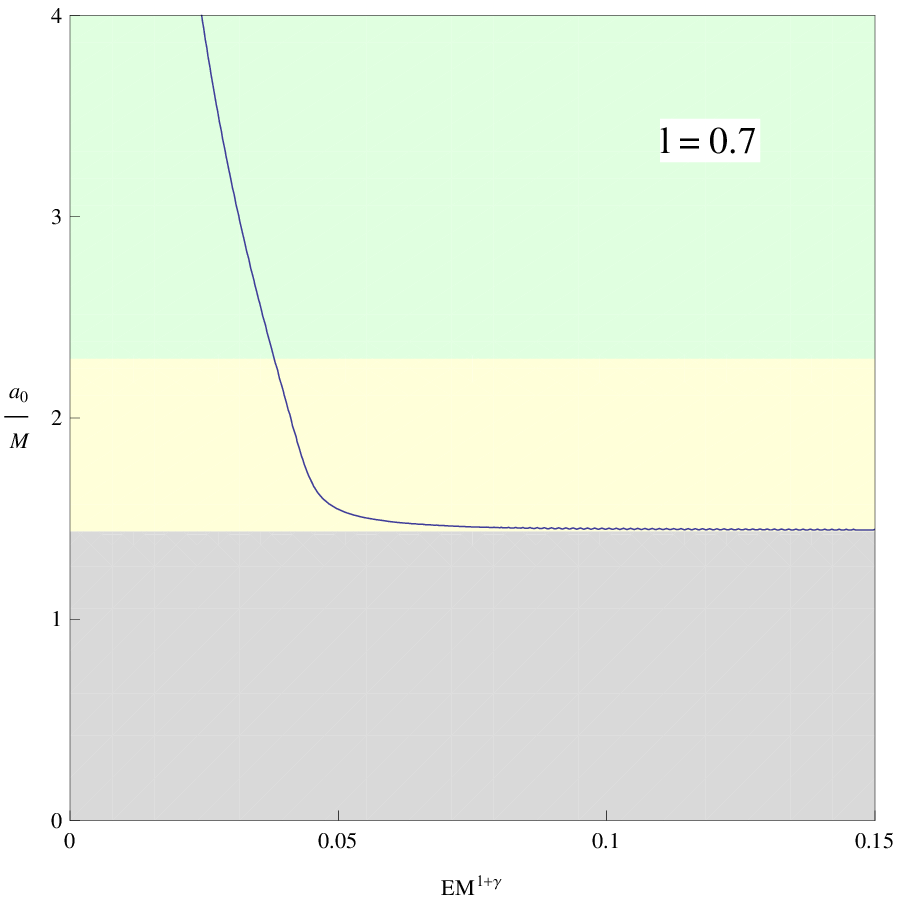,width=0.55\linewidth}\epsfig{file=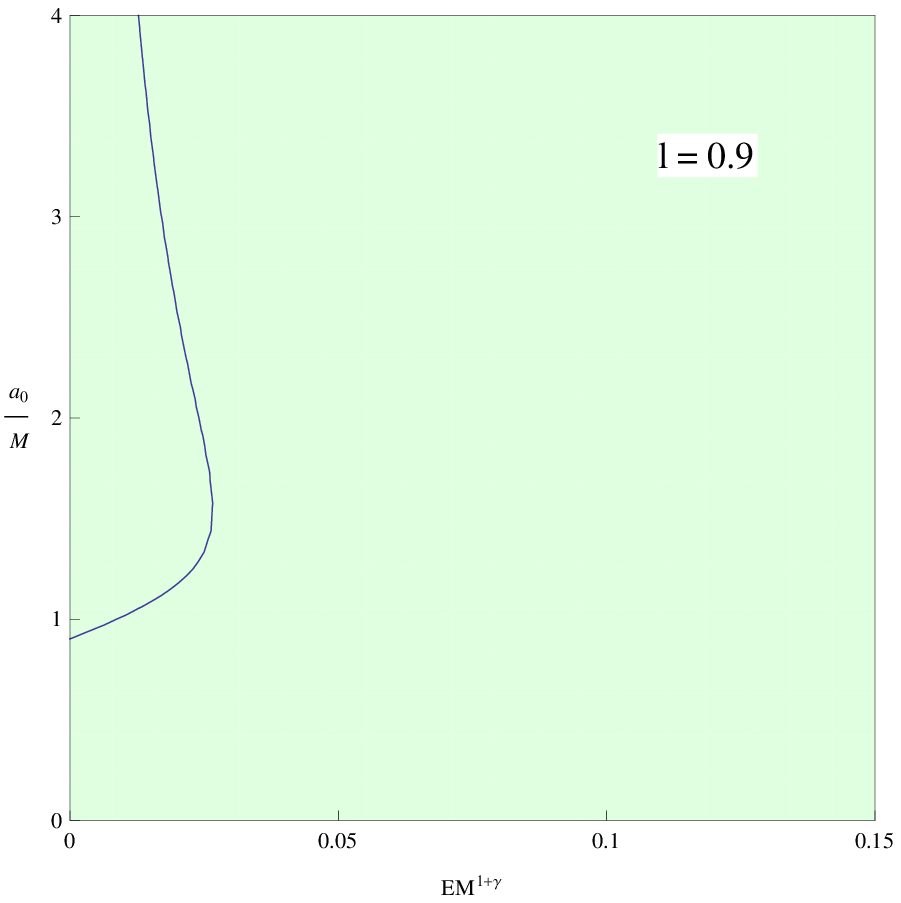,
width=0.55\linewidth}\caption{Plots for GCCG with parameters
$\gamma=0.2,~M=1,~w=-10$ and different values of $l$. Here
$EM^{(1+\gamma)}$ and $\frac{a_{0}}{M}$ are labeled along abscissa
and ordinate, respectively.}
\end{figure}
\begin{figure}\center
\epsfig{file=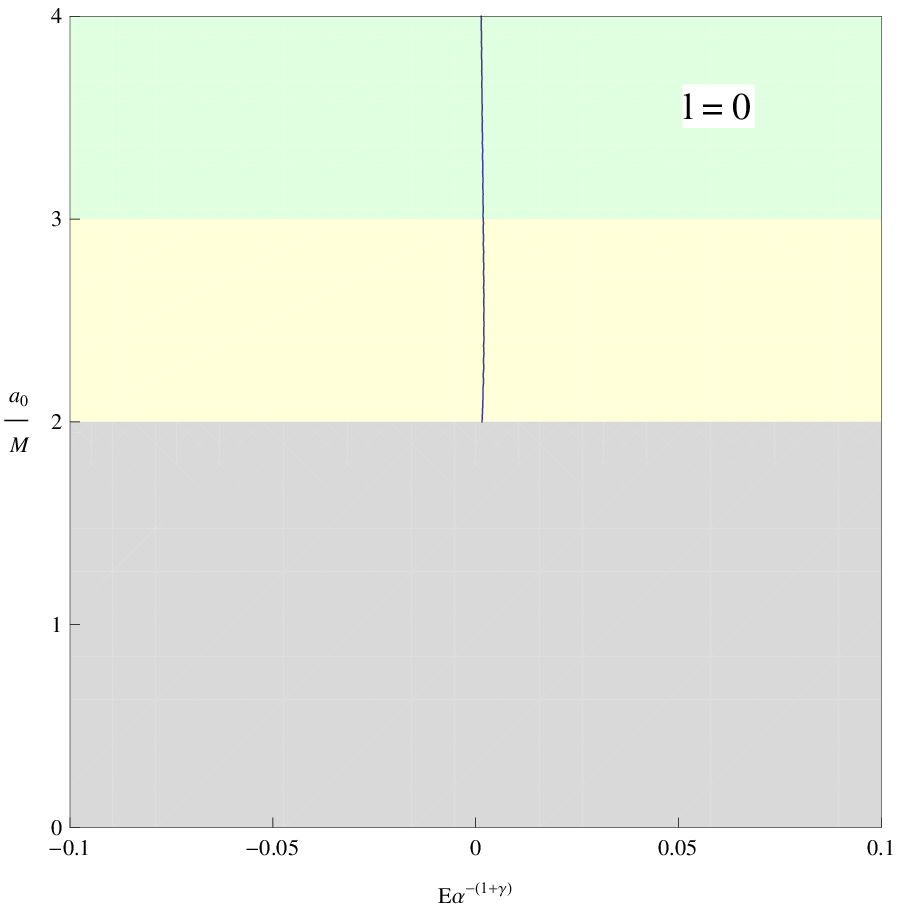,width=0.55\linewidth}\epsfig{file=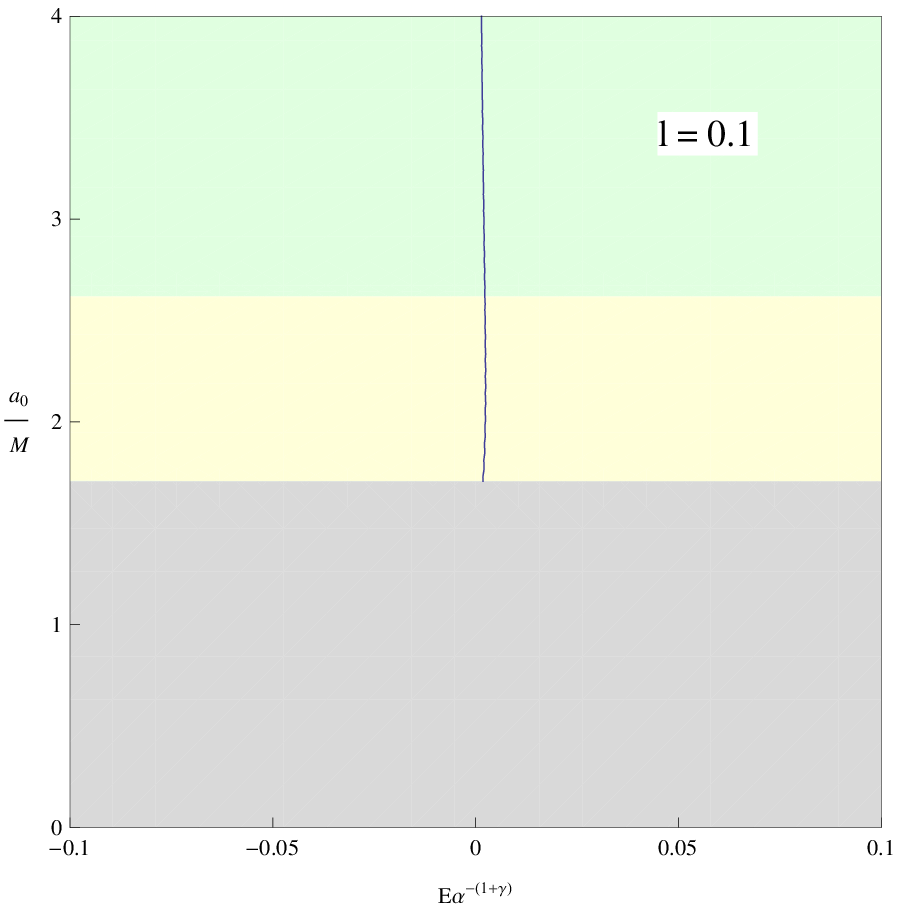,
width=0.55\linewidth}\\
\epsfig{file=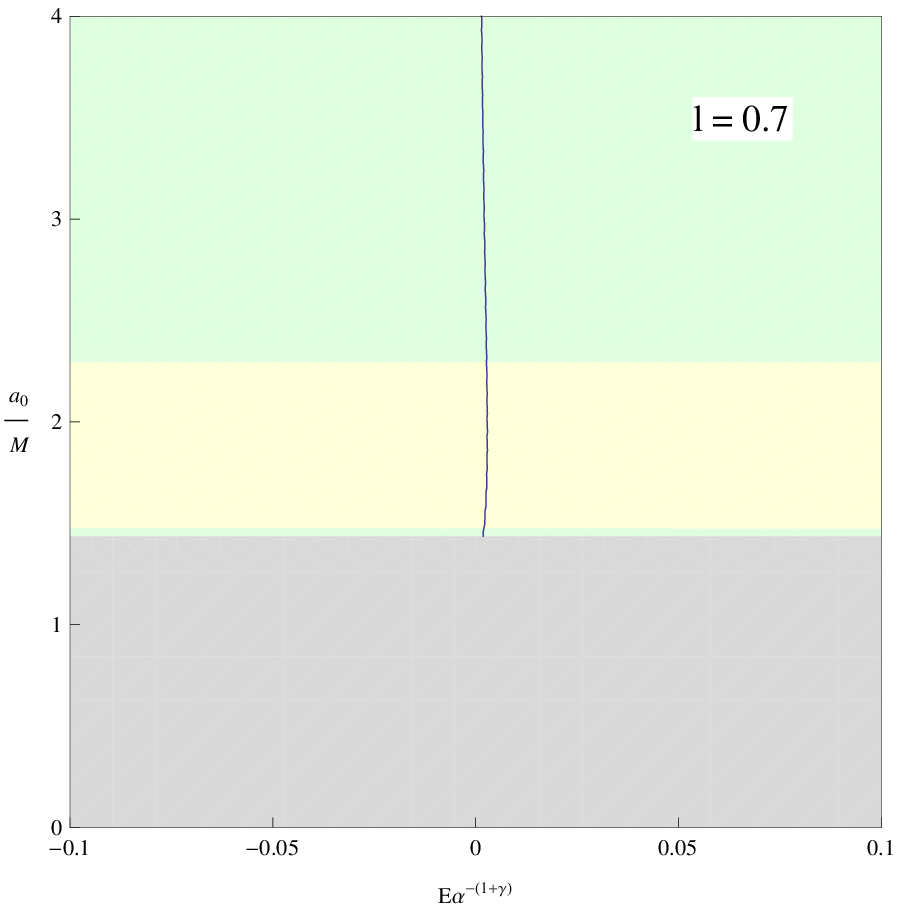,width=0.55\linewidth}\epsfig{file=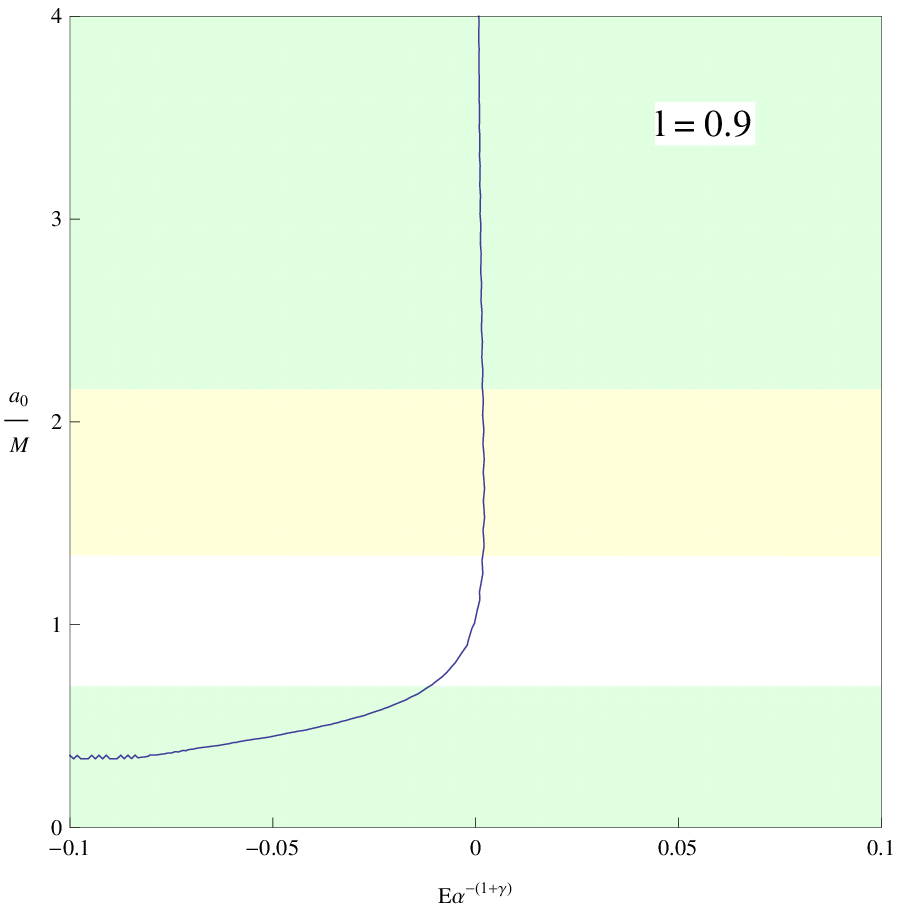,
width=0.55\linewidth}\caption{Plots for GCCG with
$\gamma=1,~M=1,~w=-10$ and different values of $l$.}
\end{figure}

\subsection{Modified Cosmic Chaplygin Gas}

Here we assume MCCG model for exotic matter for which EoS is given
by
\begin{equation}\label{31}
p=A\sigma-\frac{1}{\sigma^\gamma}\left[E+(\sigma^{1+\gamma}-E)^{-w}\right].
\end{equation}
Sadeghi and Farahani \cite{25} assumed MCCG as varying by
considering $E$ as a function of scale factor $a$, while in our
case, $E$ is assumed as a constant. Substituting Eq.(\ref{12}) in
(\ref{31}), the dynamical equation (for static configuration) is
\begin{eqnarray}\nonumber
&&[a^2_{0}F'(a_{0})+2a_{0}F(a_{0})](1+2A)[2a_{0}]^\gamma-2(4\pi
a^2_{0})^{1+\gamma}[F(a_{0})]^{\frac{1-\gamma}{2}}
\\\label{32}&&\times\left[E+\left\{(2\pi
a_{0})^{-(1+\gamma)}(F(a_{0}))^{\frac{(1+\gamma)}{2}}-E\right\}^{-w}\right]
=0.
\end{eqnarray}
The first derivative of EoS with respect to $a$ yields
\begin{align}\label{33}
\sigma'(a)+2p'(a)=\sigma'(a)\left[1+2(1+\gamma)\{A+w(\sigma^{1+\gamma}-E)^{-(1+w)}\}-\frac{2\gamma
p(a)}{\sigma(a)}\right].
\end{align}
It is noted that $\Psi(a)=\Psi'(a)=0$ at $a=a_{0}$, while the second
derivative of $\Psi(a)$ through Eq.(\ref{33}) takes the form
\begin{eqnarray}\nonumber
\Psi''(a_{0})&=&F''(a_{0})+\frac{(\gamma-1)[F'(a_{0})]^2}{2F(a_{0})}
+\frac{F'(a_{0})}{a_{0}}\left[1+2\left\{A+w(1+\gamma)\right.\right.\\\nonumber
&\times&\left.\left.\left\{\left(\frac{\sqrt{F(a_{0})}}{2\pi
a_{0}}\right)^{1+\gamma}-E\right\}^{-(1+w)}\right\}\right]-
\frac{2F(a_{0})}{a_{0}^2}(1+\gamma)\\\label{35}
&\times&\left[1+2(1+\gamma)\left[A+ w\left
\{\left(\frac{\sqrt{F(a_{0})}}{2\pi
a_{0}}\right)^{1+\gamma}-E\right\}^{-(1+w)}\right]\right].
\end{eqnarray}
For Hayward wormhole static solutions, Eq.(\ref{32}) turns out to be
\begin{align}\nonumber
&a_{0}^3(a_{0}^3+Ml^2-4Ma_{0}^2)+4M^2l^2(l^2-4a_{0}^2)+2A-(2\pi
a_{0})^{1+\gamma}\\\nonumber&\times(a_{0}^3+2Ml^2)^{\frac{3+\gamma}{2}}
(a_{0}^3+2Ml^2-2Ma_{0}^2)^{\frac{-(1+\gamma)}{2}}\left[E+\{(2\pi
a_{0})^{-(1+\gamma)}\right.\\\label{36}&\times\left.
(a_{0}^3+2Ml^2)^{\frac{-(1+\gamma)}{2}}(a_{0}^3+2Ml^2-2Ma_{0}^2)
^{\frac{(1+\gamma)}{2}}-E\}^{-w}\right]=0.
\end{align}
The results in Figures \textbf{7}-\textbf{8} show that for
$l=0,0.1,0.7$, one stable as well as unstable solution exist for
$\gamma=0.2,0.6$. We analyze that two stable and two unstable
regions appear for $l=0.9$. There exists a non-physical region for
$0<a_{0}<2$ again showing the event horizon which continues to
decrease and eventually vanishes for $l=0.9$ making a traversable
wormhole. For $l=0.9$, we find a traversable wormhole solution with
fluctuating behavior of wormhole throat which shows a stable
wormhole solution with throat expansion. It is noted that the
stability region increases by increasing values of $l$.
\begin{figure}\center
\epsfig{file=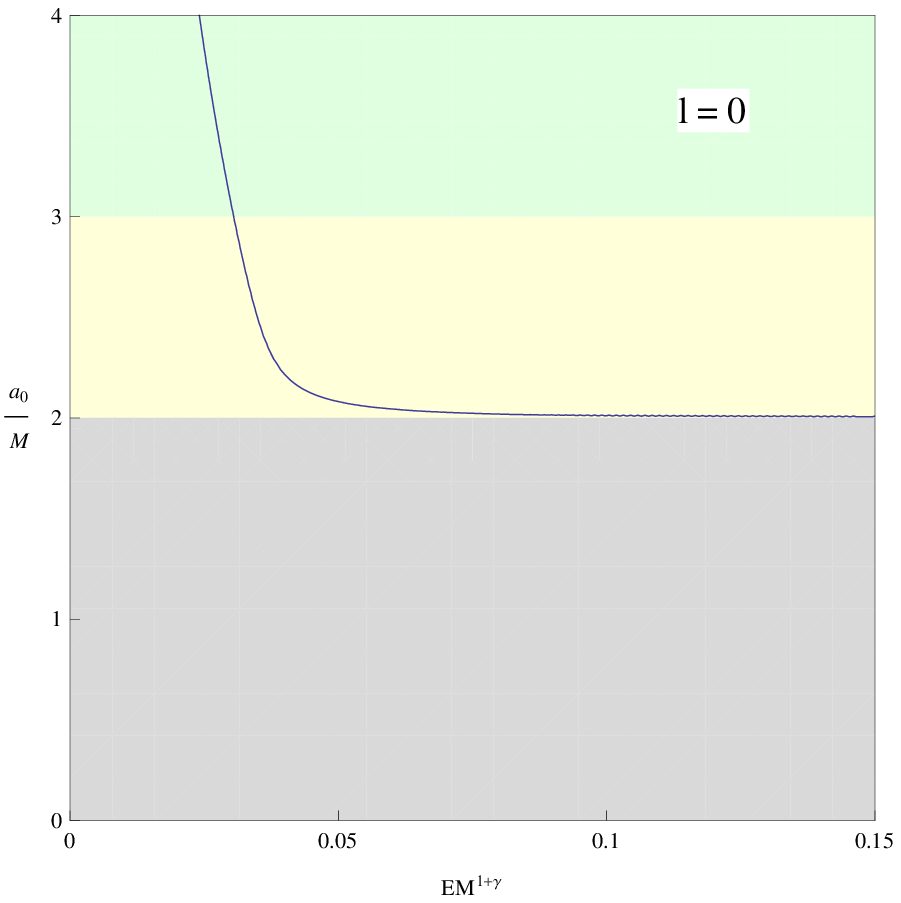,width=0.55\linewidth}\epsfig{file=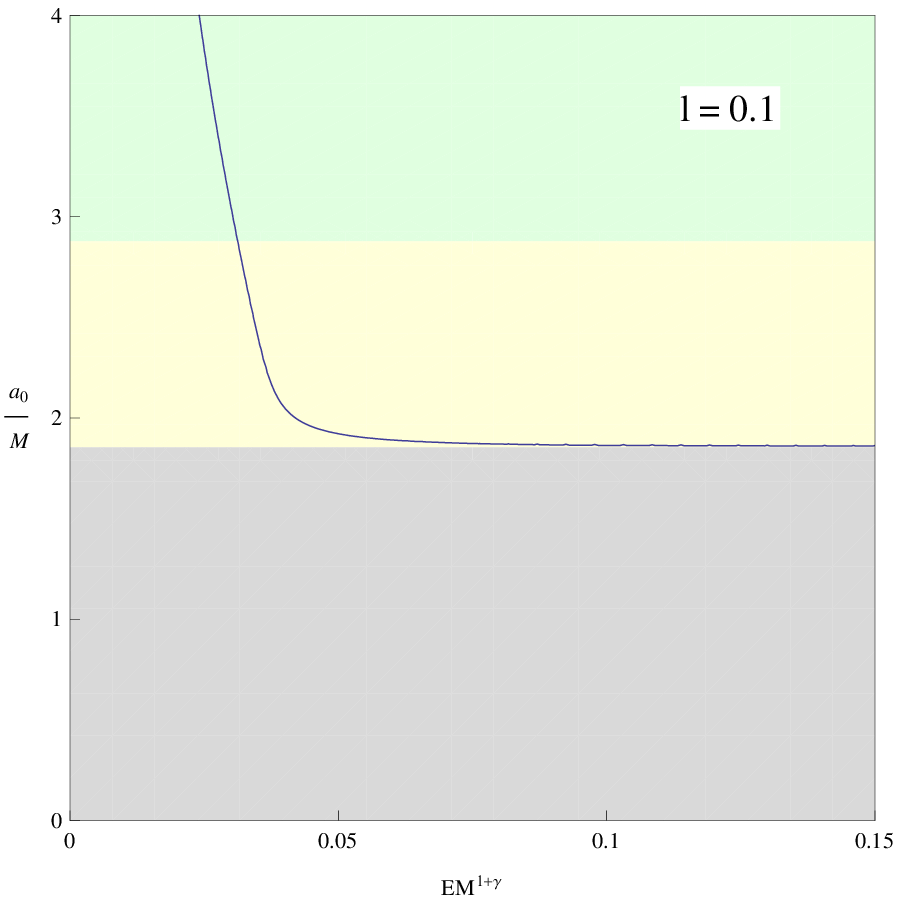,
width=0.55\linewidth}\\
\epsfig{file=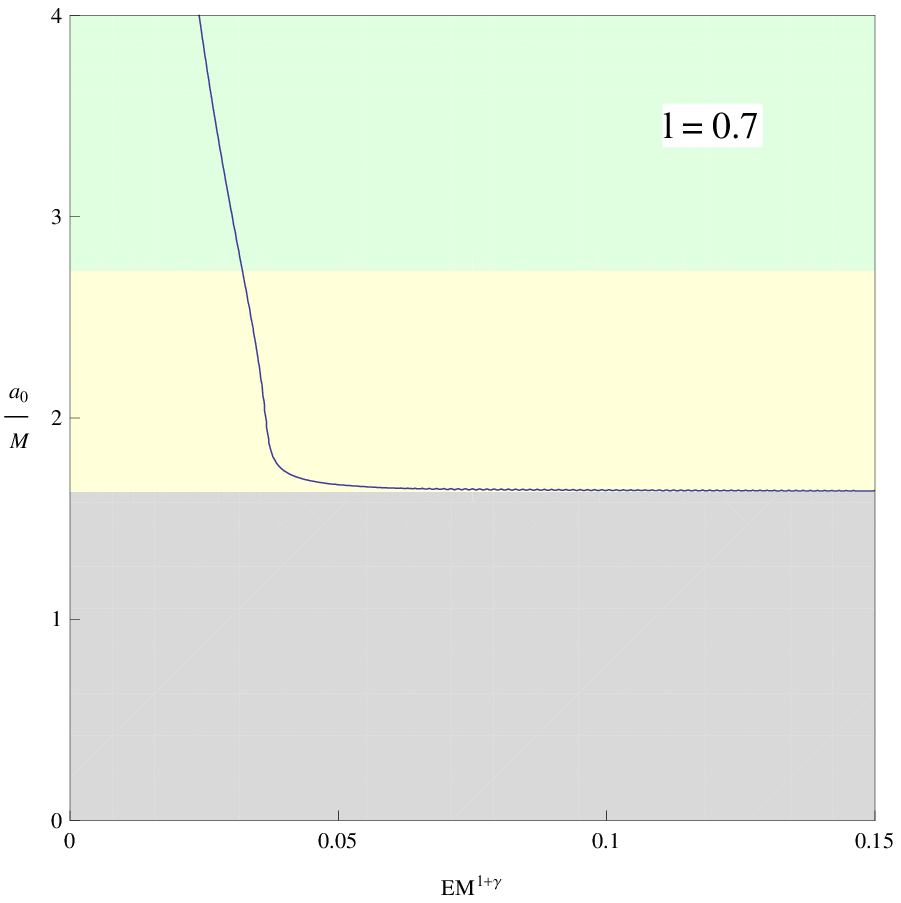,width=0.55\linewidth}\epsfig{file=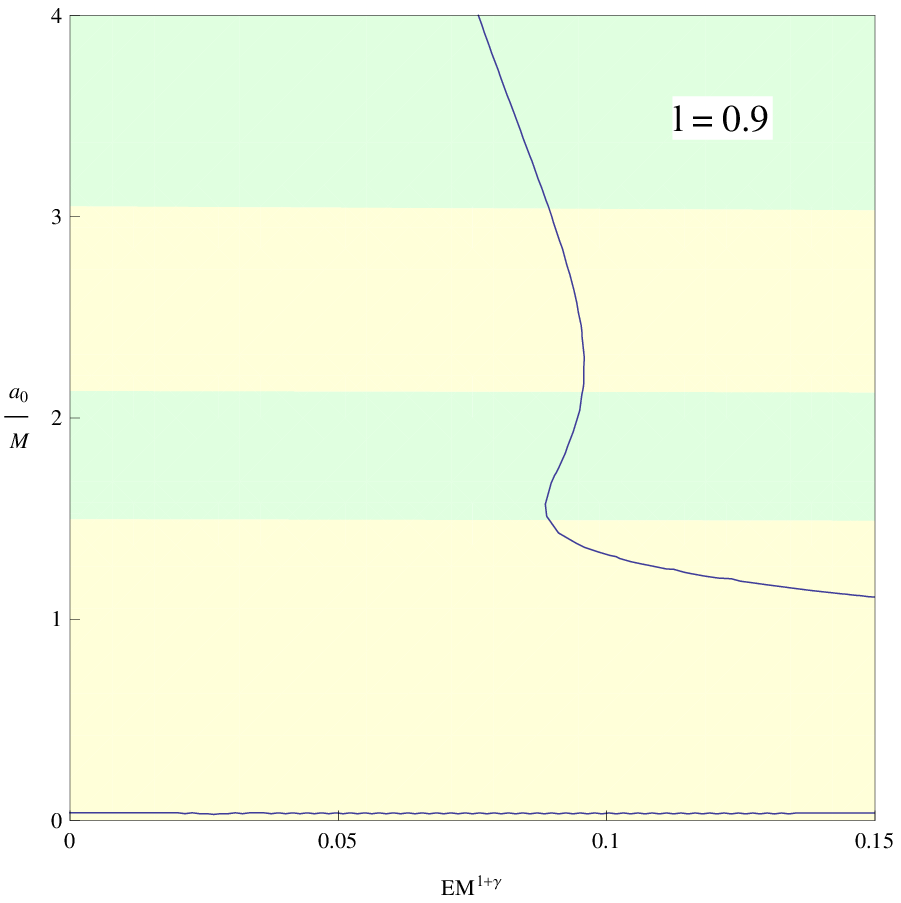,
width=0.55\linewidth}\caption{Plots for MCCG with
$\gamma=0.2,~M=1,~A=1$ and $w=-10$ with different values of $l$.
Here $EM^{(1+\gamma)}$ and $\frac{a_{0}}{M}$ are labeled along
abscissa and ordinate, respectively.}
\end{figure}
\begin{figure}\center
\epsfig{file=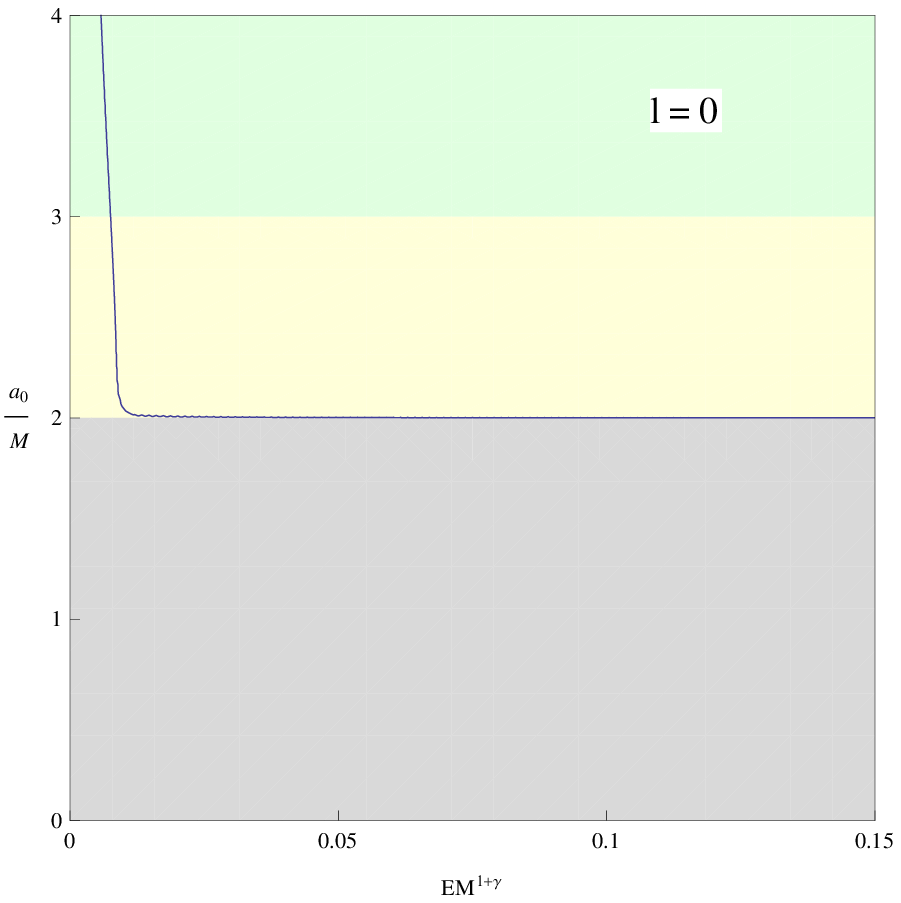,width=0.55\linewidth}\epsfig{file=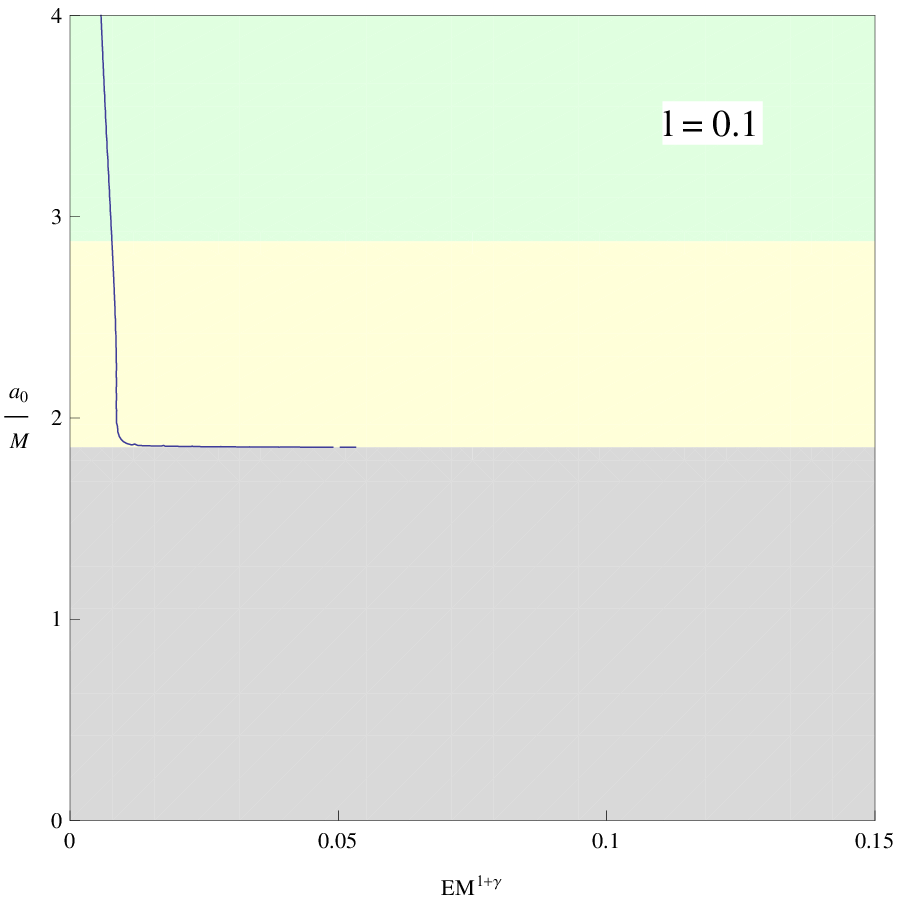,
width=0.55\linewidth}\\
\epsfig{file=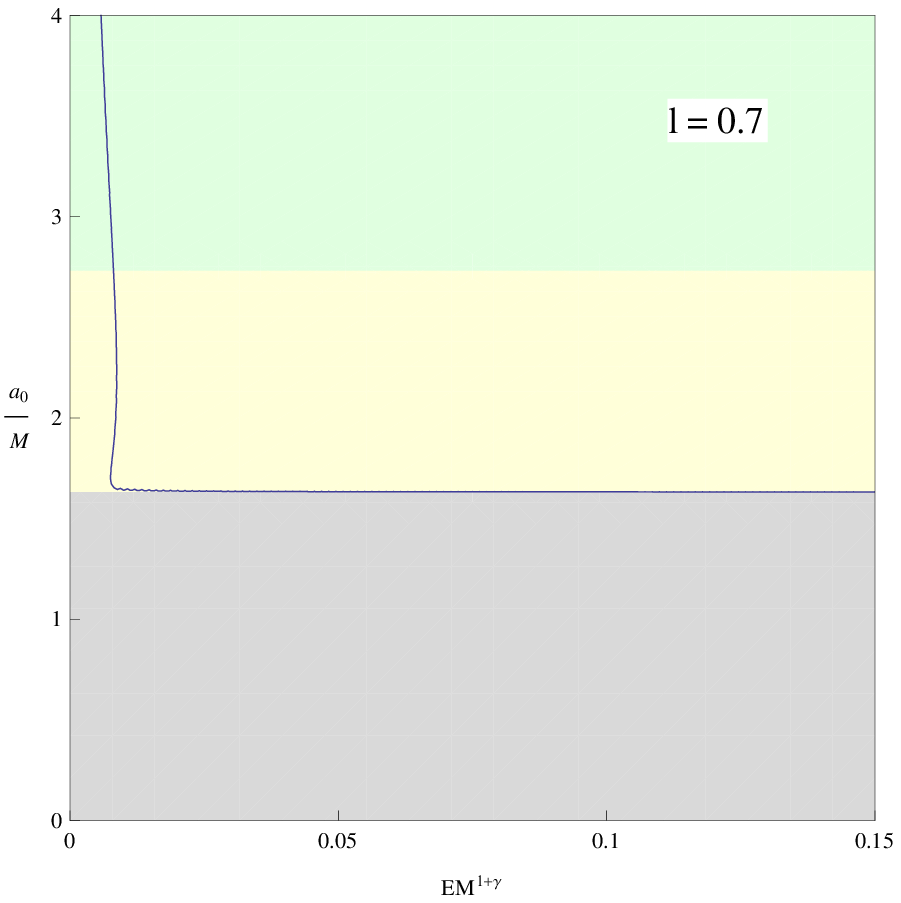,width=0.55\linewidth}\epsfig{file=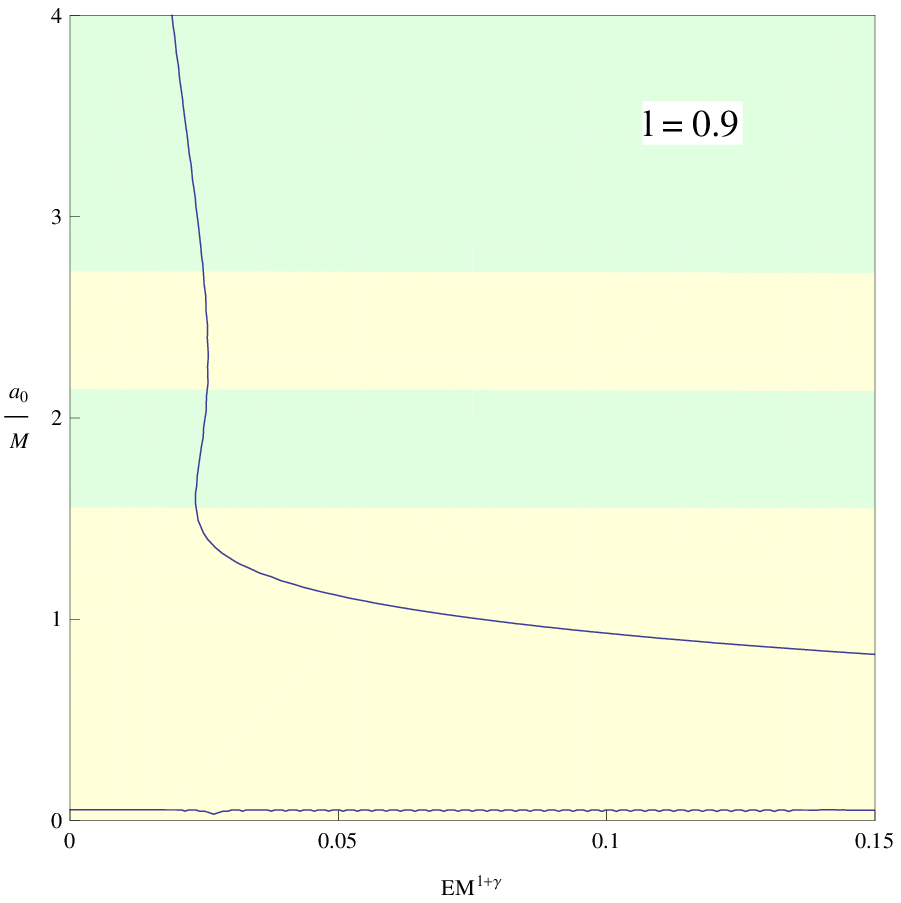,
width=0.55\linewidth}\caption{Plots corresponding to MCCG with
parameters $\gamma=0.6,~M=1,~A=1,~w=-10$ and different values of
$l$.}
\end{figure}

\section{Concluding Remarks}

In this paper, we have constructed regular Hayward thin-shell
wormholes by implementing the Visser's cut and paste technique and
analyzed their stability by incorporating the effects of increasing
values of Hayward parameter. The surface stresses have been found by
using Lanczos equations. The sum of surface stresses of matter
indicates the violation of NEC which is a fundamental ingredient in
wormhole physics leading to the existence of exotic matter. It is
found that the wormhole has attractive or repulsive characteristics
corresponding to $a^r>0$ and $a^r<0$, respectively. For a convenient
trip through wormhole, an observer should not be dragged away by
enormous tidal forces which requires that the accelerations felt by
observer must not exceed the Earth's acceleration. The construction
of viable thin-shell wormholes depends on total amount of exotic
matter confined within the shell. Here we have quantified the total
amount of exotic matter by the volume integral theorem which is
consistent with the fact that a small quantity of exotic matter is
needed to support the wormhole. We have obtained a dynamical
equation which determines possible throat radii for the static
wormhole configurations.

It is found that one may construct a traversable wormhole
theoretically with arbitrarily small amount of fluids describing
cosmic expansion. In this context, we have analyzed stability of the
regular Hayward thin-shell wormholes by taking VDW quintessence,
GCCG and MCCG models at the wormhole throat. Firstly, we have
investigated the possibility of stable traversable wormhole
solutions using VDW quintessence EoS which describes cosmic
expansion without the presence of exotic fluids. We have analyzed
both stable and unstable wormhole configurations for small values of
EoS parameter $\gamma$. The graphical analysis shows a non-physical
region (grey zone) for $0<a_{0}<2.8$ (Figures
\textbf{3}-\textbf{4}). In this region, the stress-energy tensor may
vanish leading to an event horizon and the wormhole becomes
non-traversable producing a black hole. The non-physical region in
the wormhole configuration decreases gradually and vanishes for
$l=0.9$. In this case, we find unstable wormhole solution as
$B\alpha^{-(1+\gamma)}$ approaches to its maximum value. We have
also examined stability of Hayward thin-shell wormholes for $\gamma
\in [1,\infty)$ and found only stable solutions with $l=0,0.1,0.7$
which show expansion of the wormhole throat. It is worth mentioning
here that regular Hayward thin-shell wormholes can be made
traversable as well as stable by tuning the Hayward parameter to its
large value. Also, it is noted that VDW quintessence fluid minimizes
the usage of exotic matter and $\gamma=1$ is the best fitted value
which induces only stable solutions.

In case of GCCG, we have analyzed fluctuating (stable and unstable)
solutions for $l=0,0.1,0.7$ and $\gamma=0.2,1$. It is found that
unstable solution exists for small values of throat radius $a_{0}$,
which becomes stable when the throat radius expands. There exists a
non-physical region for $0<a_{0}<2$ which diminishes for $l=0.9$
making a stable traversable wormhole. For $\gamma=1$, we have
analyzed stable regular Hayward thin-shell wormhole for throat
radius $a_{0}>2$ which undergoes expansion. We have found that only
stable wormhole configurations exist for the Hayward parameter
$l=0.9$. Finally, for MCCG, we have found one stable and one
unstable region for $l=0,0.1,0.7$, while these regions become double
(two stable and two unstable) for $l=0.9$ (Figures
\textbf{7}-\textbf{8}). There exists a non-physical region for
$0<a_{0}<2$ again showing the event horizon which eventually
vanishes for $l=0.9$ making a traversable wormhole. It is noted that
the stability region increases by increasing values of $l$. It was
found that small radial perturbations yield no stable solutions for
the regular Hayward thin-shell wormholes \cite{22}. We conclude that
stable regular Hayward wormhole solutions are possible against
radial perturbations for VDW quintessence, GCCG and MCCG models. The
trivial case $l=0$ corresponds to the Schwarzschild thin-shell
wormhole. It is worth mentioning here that the Hayward parameter
increases stable regions for regular Hayward thin-shell wormholes.\\\\
\textbf{The authors have no conflict of interest.}

\end{document}